\documentclass{mn2e}
\usepackage{epsf,times}
\newcommand{\etal}{{et al}\/.}
\begin{document}
\title[Infrared jet in Cen A]{The infrared jet in Centaurus A:
  multiwavelength constraints on emission mechanisms and particle acceleration}
\author[M.J.~Hardcastle \etal]{M.J.\ Hardcastle$^1$, R.P. Kraft$^2$ and D.M. Worrall$^3$\\$^1$ School of Physics,
  Astronomy and Mathematics, University of
Hertfordshire, College Lane, Hatfield, Hertfordshire AL10 9AB\\
$^2$ Harvard-Smithsonian Center for Astrophysics, 60 Garden Street, Cambridge, MA~02138, USA\\
$^3$ Department of Physics, University of Bristol, Tyndall Avenue,
Bristol BS8 1TL}
\maketitle
\begin{abstract}
We report on {\it Spitzer} and Gemini observations of the jet of
Centaurus A in the infrared, which we combine with radio, ultraviolet
and X-ray data. {\it Spitzer} detects jet emission from about 2 arcmin
from the nucleus, becoming particularly bright after the jet flare
point at $\sim 3.4$ arcmin. Where X-ray and infrared emission are seen
together the broad-band data strongly support a synchrotron origin for
the X-rays. The jet flare point is marked by a broad,
diffuse region of X-rays which may be associated with a shock: we
discuss possible physical mechanisms for this. The infrared jet
persists after the flare point region although X-ray emission is
absent; it is plausible that here we are seeing the effects of
particle acceleration followed by downstream advection with
synchrotron losses. Gemini data probe the inner regions of the jet,
putting limits on the mid-infrared flux of jet knots.
\end{abstract}


\section{Introduction}

Centaurus A is the closest radio galaxy (we adopt $D = 3.4$ Mpc,
Israel 1998) as well as the closest active galaxy and large
elliptical. Its proximity makes it a vital laboratory for AGN
studies of all kinds. Its nucleus shows both heavily obscured and
unobscured X-ray components (Evans \etal\ 2004), making it one of the
few low-power radio galaxies to show X-ray evidence for the obscuring
torus of canonical unification models (Evans \etal\ 2006). Its X-ray
jet, one of the first to be discovered (Schreier \etal\ 1979) has been
studied in great detail with {\it Chandra} (Kraft \etal\ 2000, 2002;
Hardcastle \etal\ 2003, hereafter H03; Kataoka \etal\ 2006). Because
of the high spatial resolution (1 arcsec = 17 pc) compared to that
available for more typical X-ray jets in more powerful FRI sources,
observations of the jet provide evidence for both localized and
diffuse particle acceleration processes. Finally, Cen A's SW radio lobe, in
expanding through the ISM of the host galaxy, drives what is currently
the only clear example of a high Mach number, attached bow shock to be
observed in X-rays around a radio galaxy (Kraft \etal\ 2003).

Cen A's main disadvantage as a subject for broad-band studies is the
strong dust lane, which obscures the inner regions of the source in
the optical to ultraviolet. Partly as a result of this, and partly
because the jet is relatively weak compared to the emission from
stars, there have until recently been no clear detections of synchrotron
emission from the jet at frequencies between radio and X-ray, although
several claims of optical and infrared emission that may be related to
the jet or material around it have been made (e.g.\ Brodie \etal\
1983, Joy \etal\ 1991, Leeuw \etal\ 2002). This contrasts with the
situation in other well-studied low-power jets, such as M87 (e.g.
Perlman \etal\ 2001) or 3C\,66B (Hardcastle \etal\ 2001) in which data
in the infrared, optical and ultraviolet support a synchrotron origin
and provide constraints on particle acceleration and
energy-production processes.

However, recent {\it Spitzer} observations have detected the jet of
Cen A in the mid-infrared (Brookes \etal\ 2006), while {\it GALEX}
detects it in the ultraviolet (Neff \etal\ 2003). In this letter we
combine these data with new and archival radio
and X-ray observations and discuss their implications for particle
acceleration processes. In addition, we use observations of the
nuclear regions with Gemini at 10 $\mu$m to place constraints on the
properties of the nucleus and the inner jet.

Except where otherwise stated, spectral indices $\alpha$ are the energy
indices, defined in the sense that flux $\propto \nu^{-\alpha}$.
\section{Observations}

\subsection{{\it Spitzer} data}

The {\it Spitzer} data we used, described in more detail by Brookes
\etal\ (2006), are taken from the public archive, and
consist of two datasets, a set of IRAC observations taken on 2004 Feb
10 and a set of MIPS observations taken on 2004 Aug 06. Several sets
of MIPS observations are available in the archive: the one we chose to
use (AOR 4940288) covers a wide area around the centre of Cen A.
The data used were the
Post-Basic Calibrated Data (PBCD) files available from the archive.
These include an automated, flux-calibrated mosaic (`MAIC' file) of
the numerous individual maps that go to make up an observation. The
PBCD files are stated in the instrument data
handbooks (http://ssc.spitzer.caltech.edu/irac/dh/ ;
http://ssc.spitzer.caltech.edu/mips/dh/) to be suitable at the time of
writing for basic scientific analysis for all IRAC channels and for
the 24-$\mu$m channel (channel 1) of the MIPS data. The low angular
resolution and calibration issues of the longer-wavelength MIPS
channels meant that these were not suitable for our analysis in any
case. IRAC channel 1 (3.6 $\mu$m) was too dominated by starlight from
the host galaxy and from foreground objects to be useful in our
analysis. Accordingly, the data we use are from the three remaining
IRAC channels, at 4.5, 5.8 and 8.0$\mu$m, and from MIPS at 24 $\mu$m;
Brookes \etal\ (2006) show a selection of images in these bands. We
carried out aperture photometry, using a local background and
excluding point sources from source and background regions, to measure
flux densities from components of the jet. Our photometry is
consistent with the independent analysis of Brookes \etal

\subsection{Gemini data}

The nuclear regions of Cen A were observed with the T-ReCS instrument
on Gemini South at $N$-band (10 $\mu$m) on 2004 Mar 06 and 2004 Mar 11-12. We
obtained these observations in an attempt to detect the bright radio
and X-ray components of the inner jet, and so the T-ReCS field of view
($28.8 \times 21.6$ arcsec) was aligned along the jet,
with the active nucleus in one corner. The standard nod and chop mode
was used for background subtraction, and the baseline calibration was
used for photometry and point-spread function (PSF) determination,
using observations of standard stars. In total the on-source exposure
time was around 2.1 h.

\subsection{VLA data}

The 8.4-GHz VLA data that we have described in earlier papers (Kraft
\etal\ 2002, H03) were not ideal for comparison with the large-scale
{\it Spitzer} jet because of the VLA's small primary beam at this
wavelength. We therefore re-reduced the data described by Clarke
\etal\ (1992) at 1.5 and 4.9 GHz. These are well matched to the
angular scales and resolution of the {\it Spitzer} data. For
smaller-scale mapping we used our existing 8.4-GHz data. VLA data from
different configurations were calibrated and combined within AIPS, and
a primary beam correction was applied to all images.

\subsection{{\it GALEX} data}

The {\it GALEX} data we use were taken from the archive
(http://galex.stsci.edu/GR1/) and were derived from observations made
on 2003 Jun 07 as part of the Nearby Galaxies Survey, as reported by
Neff \etal\ (2003); Brookes \etal\ (2006) show an image. Two broad
bandpasses are available, with mean wavelengths of 153 and 231 nm. We
use the background-subtracted intensity map, with units of (corrected)
counts s$^{-1}$, for our measurements. Photometry was carried out in
the same way as for the {\it Spitzer} data, using ground-based
calibration
(http://galexgi.gsfc.nasa.gov/Documents/ERO\_data\_description\_2.htm),
correcting for a Galactic $E(B-V)$ of 0.114 mag using the extinction
curves of Cardelli, Clayton \& Mathis (1989), which give correction
factors of 0.94 mag at both mean wavelengths. Since the photometric
zero point is not yet well defined and the extinction correction
varies significantly over the bandpasses, there are potentially large
systematic errors in the conversion between GALEX count rate and flux
density.

\subsection{{\it Chandra} data}

For the high-energy constraints on the spectrum of the jet we used two
{\it Chandra} datasets taken using the ACIS-S instrument: the
observation taken on 2002 Sep 03 (obsid 2978) which was taken as part
of the HRC guaranteed time programme, and the observation taken on
2003 Sep 14 (obsid 3965) which was taken by us in guest observer time.
These two observations are well matched in position on the instrument
and roll angle. The data were reprocessed and filtered using {\sc
ciao} 3.2.2 and {\sc caldb} 3.1 (applying new bad pixel files,
removing afterglow detection, and removing the 0.5-arcsec pixel
randomization) and were both aligned to the radio core position. After
filtering they had livetimes of 44592 and 49518 s respectively, giving
a total effective on-source time of 94.1 ks. Spectra were extracted
from regions matched to those used at other wavelengths, with local
background subtraction, using the {\it acisspec} tool within {\sc
ciao} and appropriate response matrices were generated with {\it
mkacisrmf}. Spectral fitting was done within {\sc xspec} 11.3.

\section{Results}

\subsection{The large-scale jet}

\begin{table*}
\caption{Flux densities from regions of the large-scale jet: see Fig.\
  \ref{zoom} for regions. Errors
  are nominal 3 per cent calibration errors (radio), errors based
  on calibration and background uncertainties (infrared and ultraviolet) or statistical errors (X-ray).}
\begin{tabular}{lrrrrrrrrrr}
\hline
Region&\multicolumn{8}{c}{Flux density}\\
&1.4 GHz&4.9 GHz&8.4 GHz&24 $\mu$m&8.0 $\mu$m&5.4 $\mu$m&4.5
$\mu$m&231 nm&153 nm&1
keV\\
&(Jy)&(Jy)&(Jy)&(mJy)&(mJy)&(mJy)&(mJy)&($\mu$Jy)&($\mu$Jy)&(nJy)\\
\hline
Inner&$3.25 \pm 0.10$&$1.67 \pm 0.05$&$1.0 \pm 0.03$&$9.0 \pm
1.8$&--&$2.4\pm 0.7$&--&$80 \pm 12$&--&$27.2 \pm 0.3$\\
Middle&$9.73 \pm 0.30$&$5.21 \pm 0.16$&$3.79 \pm 0.11$&$15.9 \pm
3.1$&$6.0 \pm 1.8$&$4.9 \pm 1.4$&$3.8 \pm 1.1$&$180 \pm 25$&$70 \pm 18$&$16 \pm 0.5$\\
Outer&$20.2 \pm 0.6$&$10.1 \pm 0.3$&--&$24.3 \pm 4.9$&$4.2 \pm 1.3$&$5.2
\pm 1.6$&$4.0 \pm 1.2$&--&--&$<3$\\
\hline
\end{tabular}
\label{fluxes}
\end{table*}

\begin{figure}
\epsfxsize 8.9cm
\hskip -0.2cm
\epsfbox{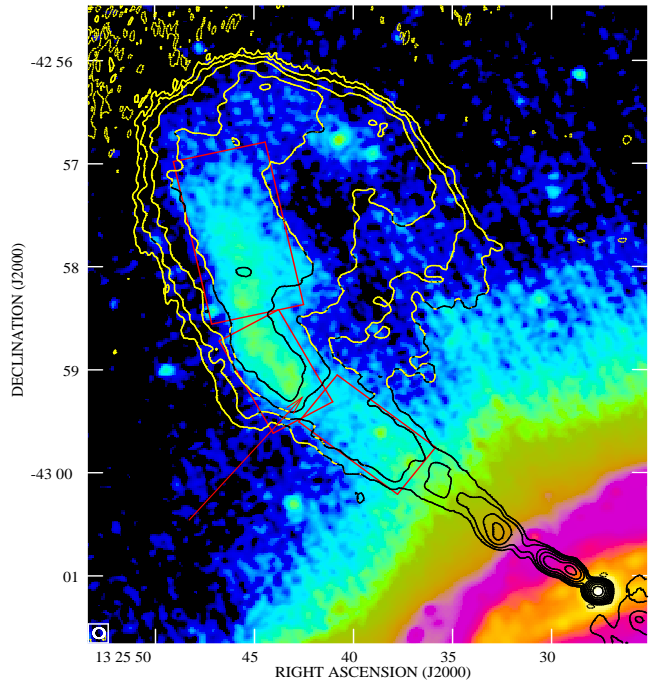}
\caption{The Cen A jet at 24 $\mu$m, using a logarithmic transfer
  function, with overlaid contours from a 6-arcsec resolution 4.9-GHz
  radio map. Black
  is the background level of 32.8 MJy sr$^{-1}$ and the peak is 724
  MJy sr$^{-1}$. The lowest contour is at 0.01 Jy beam$^{-1}$; contours
  increase by a factor 2. The flare point (peak surface brightness
  33.4 MJy sr$^{-1}$) is marked with an arrow and
  the regions used for flux measurement are shown as red boxes. }
\label{zoom}
\end{figure}

In the 24-$\mu$m data extended emission from the jet is clearly
detected from about 2 arcmin (Fig.\ \ref{zoom}) and extends at
least until the end of the clearly defined radio jet. (Two strong infrared
point sources in the lobe to the NW of the jet are probably unrelated
to it.) The brightest region of the infrared jet, and the part most
clearly detected against the higher background in the
shorter-wavelength IRAC images, occurs at a region where the radio jet
becomes abruptly brighter, and starts to bend northwards, at $\sim
3.4$ arcmin from the nucleus. Here we refer to this as the `flare
point' (not to be confused with the inner jet flare point at
around 14 arcsec from the nucleus: see H03). Ultraviolet emission is detected in
the {\it GALEX} data both from this flare point and from regions of
the jet closer to the nucleus.

\begin{figure*}
\epsfxsize 16.5cm
\epsfbox{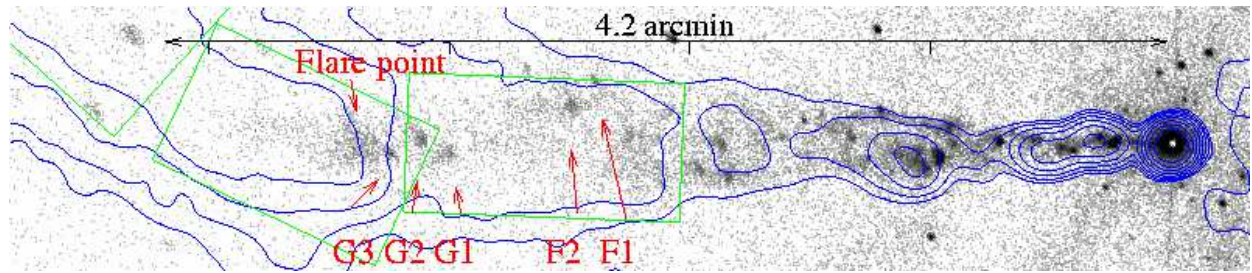}
\caption{Unsmoothed {\it Chandra} greyscale image of the jet of Cen A
  in the 0.5-5.0 keV bands. Overlaid are the contours from Fig.\
  \ref{zoom}. Pixels are the standard {\it Chandra} pixels, 0.492
  arcsec on a side. Green boxes show (from right to left) the
  boundaries of inner, middle and outer extraction regions.}
\label{chjet}
\end{figure*}

\begin{figure}
\epsfxsize 8.5cm
\epsfbox{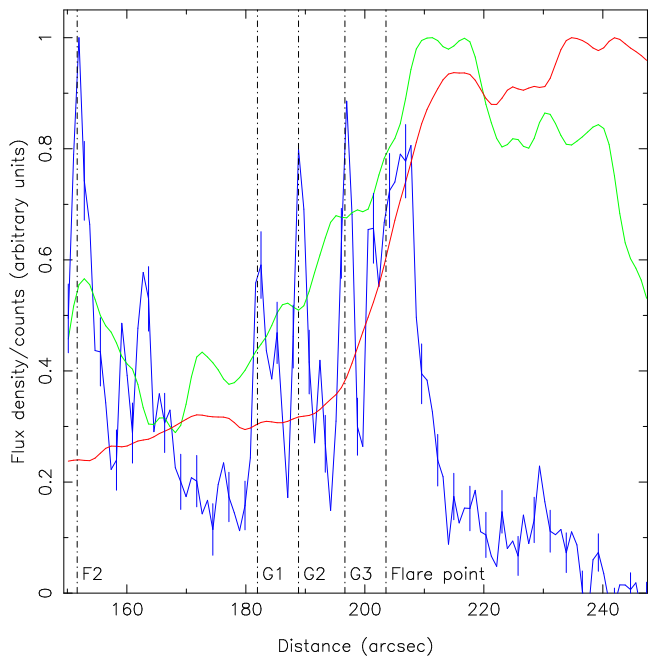}
\caption{The profile of the part of the jet near the flare point. The
  $x$-axis shows linear distance from the nucleus along the jet. Red
  indicates radio emission (from a 4.9-GHz map with $6.2 \times 2.0$
  arcsec resolution, beam elongated N-S), green indicates 24-$\mu$m infrared, and blue
  indicates 0.5-5.0 keV X-rays. The positions of the X-ray knots F2, G1, G2,
  G3 and of the flare point are marked with vertical lines. The
  extraction region for the profile was a rectangle 42 arcsec in the
  transverse direction: each point represents a 0.9-arcsec slice.
  Infrared and X-ray data were background-subtracted using adjacent
  identical regions. Representative Poisson errors are plotted on the
  X-ray points.}
\label{slice}
\end{figure}

Fig.\ \ref{chjet} shows that the flare point is marked by a
region of relatively strong X-ray emission which begins about 0.2
arcmin closer to the nucleus than the flare point and continues for
about 1 arcmin. Thereafter almost no X-ray emission is seen from the
jet, but the infrared emission continues. The X-ray region around the
flare point (which we denote region G, following the notation of
Feigelson \etal\ 1981) is resolved into several compact knots and
extended emission. The compact knots are all closer to the nucleus
than the flare point, but the brightest diffuse emission is
coincident with the flare point, although the peak radio and
infrared surface brightness (at about 3.5 arcmin) is offset from the
peak X-ray surface brightness. This is illustrated by Fig.\
\ref{slice}, which shows a profile along the jet in radio, infrared
and X-ray. The fact that the infrared surface brightness begins to
rise around knot G1 might imply that that the three knots are involved
in high-energy particle acceleration, rather than being unrelated to
the jet. There are no detected radio counterparts to these knots, but
we do not have sensitive high-resolution radio data at this distance
from the nucleus. The X-ray spectra of the knots are all well fitted
with power laws with Galactic absorption ($N_H = 7\times 10^{20}$
cm$^{-2}$) and have steep spectra (with photon indices of $1.84 \pm
0.16$, $2.01 \pm 0.16$, and $2.02 \pm 0.23$ respectively). However,
their flux densities are low ($\sim 2$ nJy each) and so they would not
contribute significantly to the flux in the infrared if these spectra
were extrapolated back to those frequencies.

\begin{figure}
\epsfxsize 8.5cm
\epsfbox{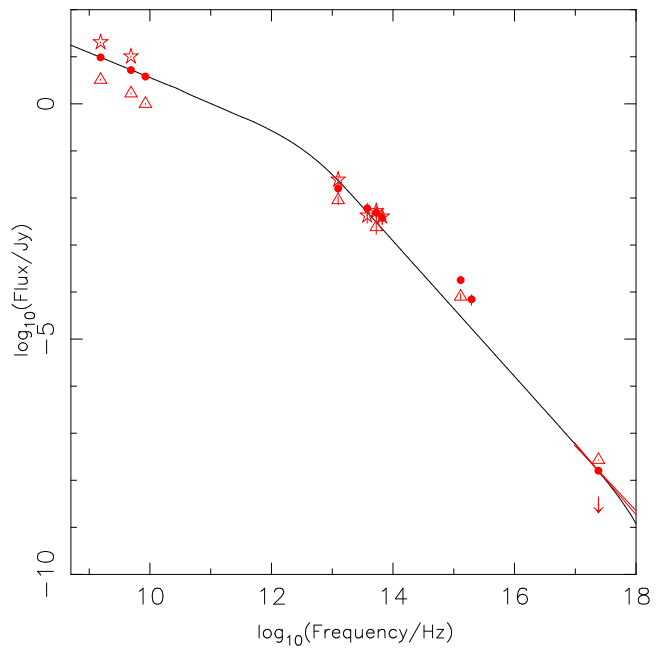}
\caption{The spectra of the inner, middle and outer regions of the jet.
  Triangles denote the inner region, filled circles the middle region,
  and stars the outer region. The solid line is the best-fitting
  reference synchrotron model to the middle region, as described in the text.
  Where error bars are not visible they are smaller than symbols. The
  X-ray upper limit is marked with an arrow.}
\label{spectra}
\end{figure}

In the absence of counterparts to the knots at other wavebands, we
exclude them in what follows, and ask the question: is the overall
spectrum of the {\it extended} emission consistent with a synchrotron
model? To investigate this, we extracted flux densities from three
matched regions of the jet around the flare point at all available
frequencies, excluding point sources and, in the case of the X-ray,
the compact knots labelled on Fig.\ \ref{chjet}, and measuring
background from adjacent off-source background regions. The X-ray
emission in
these regions is dominated by the extended emission and so the
exclusion of the knots makes little difference to our results. Fig.\
\ref{zoom} shows the extraction regions, which we call the inner,
middle and outer regions, and the results are tabulated in Table
\ref{fluxes}. The high background and low signal in the inner jet
means that we cannot measure reliable fluxes for the inner jet at 8
and 4.5 $\mu$m. For the X-ray data, we fitted power laws with Galactic
absorption to the two regions in which significant counts were
detected to determine a 1-keV flux density (finding photon indices of
$2.29 \pm 0.05$ and $2.44 \pm 0.07$ for the inner and middle regions
respectively), and the upper limit to the flux density in the outer
region was determined assuming a spectrum similar to that of the
middle jet region. At other wavelengths the flux densities were
determined by direct aperture photometry with background subtraction,
excluding point sources. To first order, the two regions with
detections in all wavebands are roughly consistent with the type of
model we have fitted elsewhere (e.g.\ Hardcastle \etal\ 2001) in which
the energy spectrum of all the electrons in the region is a broken
power law with a non-standard break connecting the radio and X-ray and
reproducing the X-ray photon indices; in this case the break must
occur at energies lower than those corresponding to the infrared
region (Fig.\ \ref{spectra}). However, in detail, the best-fitting
infrared spectral indices ($0.89 \pm 0.25$ and $0.84 \pm 0.18$ for the
inner and middle regions respectively) are somewhat flatter than would
be expected in this model, and the ultraviolet data points lie
significantly above it, closer to a linear extrapolation from the
infrared. In the middle region, in particular, where the {\it GALEX}
fluxes are probably most reliable, the data appear to require a `bump'
above the best-fitting line in which the spectrum steepens and then
flattens, which would imply a more complex electron population than
our model allows for. The outer region, in which no significant
ultraviolet or X-ray emission is detected, can also be fitted with a
simple model, but in this case either the change of the electron
spectral index, $\Delta p$, must be greater or there must be a cutoff
in the electron spectrum between the infra-red and X-ray regions; the
X-ray upper limit precludes fitting this region with a model identical
to that used in the other two regions (Fig.\ \ref{spectra}). The
best-fitting power law spectral index to the infrared data alone is
$1.13 \pm 0.19$ for the outer region, which would be consistent with a
larger $\Delta p$, though the errors are large. In such a model there
may be low-level X-ray emission from this region that would be
detectable in deeper observations.

No obvious jet-related infrared emission is seen on the counterjet
side of the source. The counterjet radio and X-ray knots discussed by
H03 are at small distances from the nucleus, where the infrared
background from the host galaxy is high.

\subsection{The small-scale jet}

The Gemini observations do not detect any component of the jet within
24 arcsec of the nucleus. Using the baseline photometric calibration
we estimate that an upper limit on any compact jet-like component is 1
mJy at 10 $\mu$m. The observations are less sensitive than would have
been predicted, presumably because of the bright emission from the
dust lane. The upper limit allows us to set a lower limit on the
spectral index between radio and infrared, given the knot radio fluxes
measured previously (H03), of $\alpha > 0.5$, consistent with the
$\alpha_{\rm RI}$ measured for the large-scale jet, $\sim 0.75$. Since
the knots in the inner region generally have steep X-ray spectral
indices, it seems likely that their spectrum, like that of the outer
jet, turns over before the infrared region. Sub-mm observations are
required to test this.

\subsection{The nucleus}

Cen A's nucleus has been observed at wavelengths around $N$-band by
several other groups (Krabbe, B\"oker \& Maiolino 2001; Karovska
\etal\ 2003; Siebenmorgen, Kr\"ugel \& Spoon 2004). We estimate the
background-subtracted 10-$\mu$m flux density of the nucleus in our
Gemini observations as 1.1 Jy, with a 10 per cent photometric
calibration error. This is significantly higher than seen in some
earlier observations, though consistent with the $1.5 \pm 0.4$ Jy
reported by Karovska \etal from data taken in 2002 May, and may
indicate variability on timescales of years. Variability has earlier
been claimed at shorter infrared wavelengths (Turner \etal\ 1992).
Comparing 825-s individual observations from 2004 Mar 06 with the
short, 43-s observation of the standard star HD 110458, we find that
the nucleus appears slightly resolved, with a Gaussian FWHM of 0.27
arcsec. Some of the apparent extension with respect to the PSF may be
the result of the longer observation, with more chop/nod cycles, used
for the nucleus, but we can conservatively say that the size of the
nucleus at 10 $\mu$m is $\la 0.27$ arcsec, or $\la 4.5$ pc, consistent
with the measurement of $0.17 \pm 0.02$ arcsec by Karovska \etal\ The
torus in Cen A must thus be compact.

\section{Discussion: the origin of the flare point}

The {\it Spitzer} observations confirm that the broad-band spectrum of
the large-scale Cen A jet can be described with a synchrotron model,
as in other FRI sources, although the detailed spectral shape almost
certainly requires a multi-component model for the synchrotron emission.
However, the infrared detection points up the importance of the region
we have called the flare point. Both the radio and infrared brighten
by a factor $\sim 3$ here (Fig.\ \ref{slice}) while there is little or
no X-ray emission after the extended component of region G. The short
synchrotron lifetime of X-ray-emitting electrons means that X-ray
emission must always be associated with high-energy particle
acceleration: but could the lack of X-rays after region G imply that
this region, which is roughly coincident with the entry of the jet
into the lobe, represents the `last gasp' of particle acceleration in
Cen A's jet? In a field strength of 3 nT, the equipartition value for
this part of the jet, the electron energy loss timescale ($E/({\rm
d}E/{\rm d}t)$) is $2 \times 10^4$ years for electrons emitting at 4.5
$\mu$m, and $4 \times 10^4$ years for electrons emitting at 24 $\mu$m.
The projected distance from the end of the X-ray emission at the flare
point to the end of the jet, where only radio and 24-$\mu$m emission
is detected, is roughly 2.5 kpc, implying a light travel time of $\sim
8 \times 10^3$ years. Thus, for particle acceleration to be absent in
this region, we require jet speeds of $(0.2/\sin\theta)c$, where
$\theta$ is the angle to the line of sight. As speeds in the inner jet
are $\ga 0.5c$, and the angle to the line of sight may be relatively
large (see discussion in H03) this is not impossible, so that it could
indeed be the case that significant high-energy particle acceleration
ceases at the flare point.

This motivates us to ask a further question: what is the nature of the
extended X-ray emission at the flare point? Most of the X-rays (Fig.\
\ref{slice}) come from a region only 10 arcsec, or 170 pc, in size.
This is still larger than the expected travel distance for
1-keV-emitting electrons (at most 100 pc) and in fact extended X-ray
emission is seen on scales up to around 30 arcsec, making it difficult
to sustain a model in which the particle acceleration here takes place
at a single point if the magnetic field strength has its equipartition
value, although only modest decreases in the magnetic field strength,
by a factor of a few, would be necessary to make a one-shot
acceleration model viable, since the loss timescale goes as
$B^{-3/2}$. More puzzling in this picture is the offset between the
peak X-ray, radio and infrared surface brightnesses seen in Fig.\
\ref{slice}. It is also not clear what the physical relationship is
between the diffuse X-ray emission at the flare point and the X-ray
knots G1--3. In fact there is a striking similarity in the X-ray,
albeit on larger scales, between the flare point and other regions in
the inner part of the jet where we see compact X-ray features followed
by diffuse X-ray emission, such as the region around knot BX2 (H03).

In H03 we argued that the knots in the inner jet were related to
shocks as a result of interactions between the jet fluid and
small-scale obstacles in the jet. The flare point is different in that
it appears to affect the whole jet. It would be tempting, since the
extended emission is associated with the entry of the jet into the
high-surface-brightness regions of the NE lobe, to suggest that we are
seeing an extended reconfinement shock, in which case the similarity
of the length of the X-ray emitting region and the width of the radio
jet implies a Mach number $\sim 2$. The roughly tapering shape of the
downstream X-ray emission is consistent with this idea (Sanders 1983)
if only the inward-propagating shock accelerates particles to high
energies. The idea that the jet decelerates rapidly and significantly
at the flare point is consistent with the observation that it becomes
both broader and brighter at this position. The details of the shock
structure would then depend on the velocity structure of the jet, and
it may be that a detailed model taking this into account could
reproduce the offsets between the emission peaks at different
frequencies. An alternative model is that the jet interacts at the
flare point with some large-scale external feature that is able to
affect the whole jet. Gopal-Krishna and Saripalli (1984) have pointed
out the coincidence between the radio and X-ray flare point and an
optical `shell', seen in deep images, which they consider to be a
remnant of a cannibalized galaxy. A series of shocks caused by such an
interaction would be equally consistent with what we observe, if they
can be distributed over the entire region of X-ray emission or, again,
if the magnetic field is sub-equipartition.

Qualitatively, the bulk deceleration at the flare point in either of
these scenarios is also consistent with the observed onset of bending
of the jet there. However, if the jet really has a high bulk speed, as
required to avoid {\it in situ} particle acceleration after the flare
point, its density must be low. For example, if we assume that the jet
is bent by the ram pressure of hot external material moving at the
sound speed, then, applying Euler's equation (e.g. Eilek \etal\ 1984)
and using the parameters of Kraft \etal\ (2003), the jet density must
be $10^{-4}$ times the external density if the speed is $0.2c$. This
is still a factor 5 above the minimum possible effective jet density
(from the minimum-energy condition, assuming that the jet is a pure
lepton/magnetic field plasma), but if any entrainment of baryonic
material takes place in the inner jet, or if there is a significant
departure from equipartion, the jet will be heavier, in which case
relativistic bulk speeds at the bend would be unrealistic. In that
situation, the post-shock jet speed would have to be slower, and some
continuing {\it in situ} particle acceleration would be required to
explain the extended infrared jet.

\section*{Acknowledgements}

We are grateful to Charles Lawrence and Mairi Brookes for discussion
of their results on Cen A prior to publication. MJH thanks the Royal
Society for a research fellowship.


\begin{thebibliography}{}
\bibitem[]{139}Brodie, J.P., K\"onigl, A., Bowyer, S., 1983, ApJ, 273, 154
\bibitem[]{140}Brookes, M.H., Lawrence, C.R., Stern, D., Werner, M., 2006, ApJ  submitted
\bibitem[]{187}Cardelli, J.A., Clayton, G.C., Mathis, J.S., 1989, ApJ, 345, 245
\bibitem[]{213}Clarke, D.A., Burns, J.O., Norman, M.L., 1992, ApJ, 395, 444
\bibitem[]{296}Eilek, J.A., Burns, J.O., O'Dea, C.P., Owen, F.N., 1984, ApJ, 278, 37
\bibitem[]{307}Evans, D.A., Kraft, R.P., Worrall, D.M., Hardcastle, M.J., Jones, C., Forman, W.R., Murray, S.S., 2004, ApJ, 612, 786
\bibitem[]{308}Evans, D.A., Worrall, D.M., Hardcastle, M.J., Kraft, R.P., Birkinshaw, M., 2006, ApJ in press, astro-ph/0512600
\bibitem[]{332}Feigelson, E.D., Schreier, E.J., Delaville, J.P., Giacconni, R., Grindlay, J.E., Lightman, A.P., 1981, ApJ, 251, 31
\bibitem[]{390}Gopal-Krishna, Saripalli, L., 1984, A\&A, 141, 61
\bibitem[]{422}Hardcastle, M.J., Birkinshaw, M., Worrall, D.M., 2001, MNRAS, 326, 1499
\bibitem[]{433}Hardcastle, M.J., Worrall, D.M., Kraft, R.P., Forman, W.R., Jones, C., Murray, S.S., 2003, ApJ, 593, 169 [H03]
\bibitem[]{498}Israel, F.P., 1998, A\&A  8 237
\bibitem[]{528}Joy, M., Harvey, P.M., Tollestrup, E.V., Sellgren, K., McGregor, P.J., Hyland, A.R., 1991, ApJ, 366, 82
\bibitem[]{539}Kataoka, J., Stawarz, L., Aharonian, F., Takahara, F., Ostrowski, M., Edwards, P.G., 2006, ApJ in press, astro-ph/0510661
\bibitem[]{570}Krabbe, A., B\"oker, T., Maiolino, R., 2001, ApJ, 557, 626
\bibitem[]{571}Kraft, R.P., et al., 2000, ApJ, 531, L9
\bibitem[]{572}Kraft, R.P., Forman, W.R., Jones, C., Murray, S.S., Hardcastle, M.J., Worrall, D.M., 2002, ApJ, 569, 54
\bibitem[]{573}Kraft, R.P., V\'azquez, S., Forman, W.R., Jones, C., Murray, S.S., Hardcastle, M.J., Worrall, D.M., Churazov, E., 2003, ApJ, 592, 129
\bibitem[]{632}Leeuw, L.L., Hawarden, T.G., Matthews, H.E., Robson, E.I., Eckart, A., 2002, ApJ, 565, 131
\bibitem[]{754}Neff, S.G., Schiminovich, D., Martin, C.D., 2003, AAS 203 96.07
\bibitem[]{827}Perlman, E.S., Biretta, J.A., Sparks, W.B., Macchetto, F.D., Leahy, J.P., 2001, ApJ, 551, 206
\bibitem[]{920}Sanders, R.H., 1983, ApJ, 266, 73
\bibitem[]{944}Schreier, E.J., Feigelson, E., Delvaille, J., Giacconi, R., Grindlay, J., Schwartz, D.A., Fabian, A.C., 1979, ApJ, 234, L39
\bibitem[]{950}Siebenmorgen, R., Kr\"ugel, E., Spoon, H.W.W., 2004, A\&A, 414, 123
\bibitem[]{1041}Turner, P.C., Forrest, W.J., Pipher, J.L., Shure, M.A., 1992, ApJ, 393, 648
\end{thebibliography}
\end{document}